# Selective Epitaxial Growth of GaAs on Ge Substrates with a SiO$_2$ Pattern

G. Brammertz[a], M. Caymax[a], Y. Mols[a], S. Degroote[a], M. Leys[a], J. Van Steenbergen[a], G. Winderickx[a], G. Borghs[a] and M. Meuris[a].

[a]Interuniversity Microelectronics Center (IMEC), Kapeldreef 75, 3001 Leuven, Belgium.

We have selectively grown thin epitaxial GaAs films on Ge substrates with the aid of a 200 nm thin SiO$_2$ mask layer. The selectively grown structures have lateral sizes ranging from 1 um width up to large areas of 1 by 1 mm$^2$. The growth is fully selective, thanks to an optimized growth procedure, consisting of a 13 nm thin nucleation layer grown at high pressure, followed by low pressure growth of GaAs. This growth procedure inhibits the nucleation of GaAs on the mask area and is a good compromise between reduction of loading effects and inhibition of anti phase domain growth in the GaAs. Nevertheless, both microscopic and macroscopic loading effects can still be observed on x-section SEM images and profilometer measurements. X-ray diffraction and low temperature photoluminescence measurements demonstrate the good microscopic characteristics of the selectively grown GaAs.

## Introduction

While conventional Si MOSFET scaling beyond the 45 nm generation encounters numerous challenges, more techniques are being developed for mobility enhancement in the channel region. With the introduction of SiGe in the channel region, the first step was taken towards the integration of higher mobility materials in traditional Si CMOS technology (1,2). In addition, with the introduction of high-k materials for gate insulation, a serious advantage of Si CMOS has vanished, the relatively simple surface passivation of Si by its natural oxide.
Ge and GaAs are intrinsically faster semiconductors and especially the implementation of GaAs NMOS ($\mu_n$ = 8000 cm$^2$V$^{-1}$sec$^{-1}$) with Ge PMOS ($\mu_p$ = 1800 cm$^2$V$^{-1}$sec$^{-1}$) on the same substrate presents serious mobility advantages over traditional Si CMOS ($\mu_n$ = 1350 cm$^2$V$^{-1}$sec$^{-1}$, $\mu_p$ = 480 cm$^2$V$^{-1}$sec$^{-1}$) (3). Figure 1 shows a schematic representation of the layout of such a Ge/GaAs co-integrated chip in a planar scalable technology. It becomes apparent that the basic need for such an implementation is selective growth of GaAs on Ge substrates.

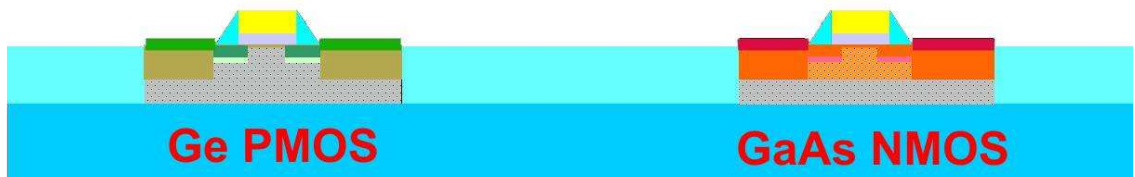

**Figure 1.** Schematic representation of a planar implementation of Ge PMOS with GaAs NMOS on the same wafer.

The feasibility of selective area growth of GaAs on Ge was already presented (4). Here we will discuss the properties of the selectively grown GaAs on Ge. First we will shortly introduce the experimental setup, mask layout and growth procedure for the selective area growth of GaAs on Ge, followed by the description of the mechanical properties of the selectively grown layer. Then we will inspect the microscopic quality of the material with low temperature photoluminescence spectroscopy and compare it to the results for plain GaAs grown on Ge substrates.

## Experimental details

The GaAs growth tool is a Thomas Swan MOCVD reactor, in which the reactants impinge on the susceptor from the top via a close-coupled showerhead. From there the gases are vertically removed from the reactor through the side of the susceptor via a quartz liner system. The total pressure in the reactor can be varied from 30 torr to atmospheric pressure. The precursor sources for the MOCVD growth are Trimethylgallium (TMGa) and Tertiarybutylarsine (TBAs).

The GaAs is grown on bulk 4" Ge (001) wafers acquired from UMICORE. For anti phase domain (APD)-free growth, the substrate needs to have a misalignment with respect to the exact [001] direction (5-7). In our case we chose for a misalignment of 6° towards the [111] direction, which allows for a sufficient density of single atomic steps on the Ge surface. On these wafers a 200 nm thick amorphous $SiO_2$ film is deposited and subsequently patterned and etched selectively in an HF solution, in order to reveal the Ge in the holes of the dielectric. The maskset chosen is a general mask used for transistor processing and contains structures as small as a micron width up to larger structures of a mm square, allowing for a very large spread in feature sizes and filling factor of the $SiO_2$ over the area of the mask. The global filling factor, which is the ratio of open area to the total area of the mask, is equal to 40%, which means that the larger part of the area is covered with $SiO_2$. Loading effects, caused by diffusion of growth species on the mask surface and desorption of growth species from the mask into the gas phase, should therefore be important, allowing a good study of the effects of selective growth.

For all growths, a substrate temperature of 660°C was adapted, as this temperature shows the best compromise between desorption of the group III precursor from the $SiO_2$, while not decomposing the group V precursor on the $SiO_2$, allowing for the best selective growth of GaAs with a $SiO_2$ mask (8). The GaAs growth time was chosen in order to obtain a 150 nm total film thickness on an unpatterned Ge substrate. For more details about the procedure for selective epitaxial growth of GaAs on Ge substrates consult reference (4). We would just like to repeat here that the growth procedure consists of the growth of a very thin (~13 nm) nucleation layer grown at higher pressure, followed by the growth of the residual layer at low pressure (30 torr) in order to reduce the loading effects due to precursor migration on and over the mask area. The higher pressure nucleation layer is necessary in order to suppress the formation of anti phase boundaries, which are electrically very active sites in the GaAs lattice, reducing the mobility of the material considerably.

## Material characterization.

Figure 2 shows optical microscope images of the selectively grown GaAs on Ge with a SiO₂ mask layer. Different structures of different sizes can be identified on the images, showing the large variation in filling factor that is present on the mask. For all feature sizes, selectivity is complete, with no GaAs nucleation on the SiO₂ mask area and full GaAs growth in the holes, where the Ge substrate is revealed.

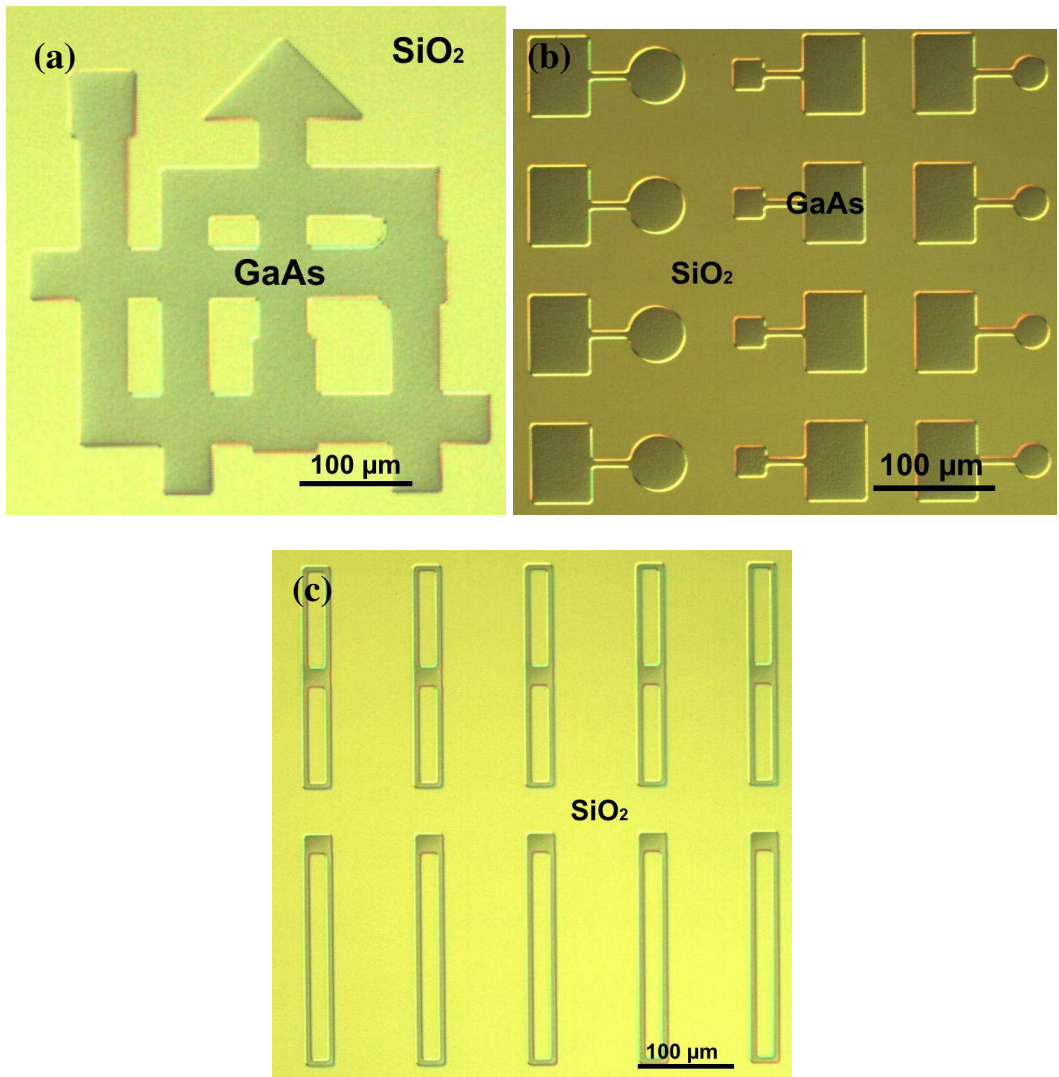

**Figure 2.** Optical microscope images of the selectively grown GaAs on Ge with a SiO2 mask layer. The figures 2 **(a)** to **(c)** show structures of different sizes, visualizing the large spread in filling factor of the mask.

Cross-section scanning electron microscopy (SEM) images reveal more detail about the structural quality of the layers. Figure 3 shows three different cross-section SEM pictures of the selectively grown GaAs. Figure 3(a) shows an overview of the cross-section of a 12 µm wide GaAs stripe. The stripe is at an angle of 45° with respect to the cleavage plane, explaining the presence of GaAs on the right hand side of the image and the longer width of the structure. A small loading effect can already be seen at the interfaces with the SiO₂, even though the overall flatness of the selectively grown layer is relatively good.

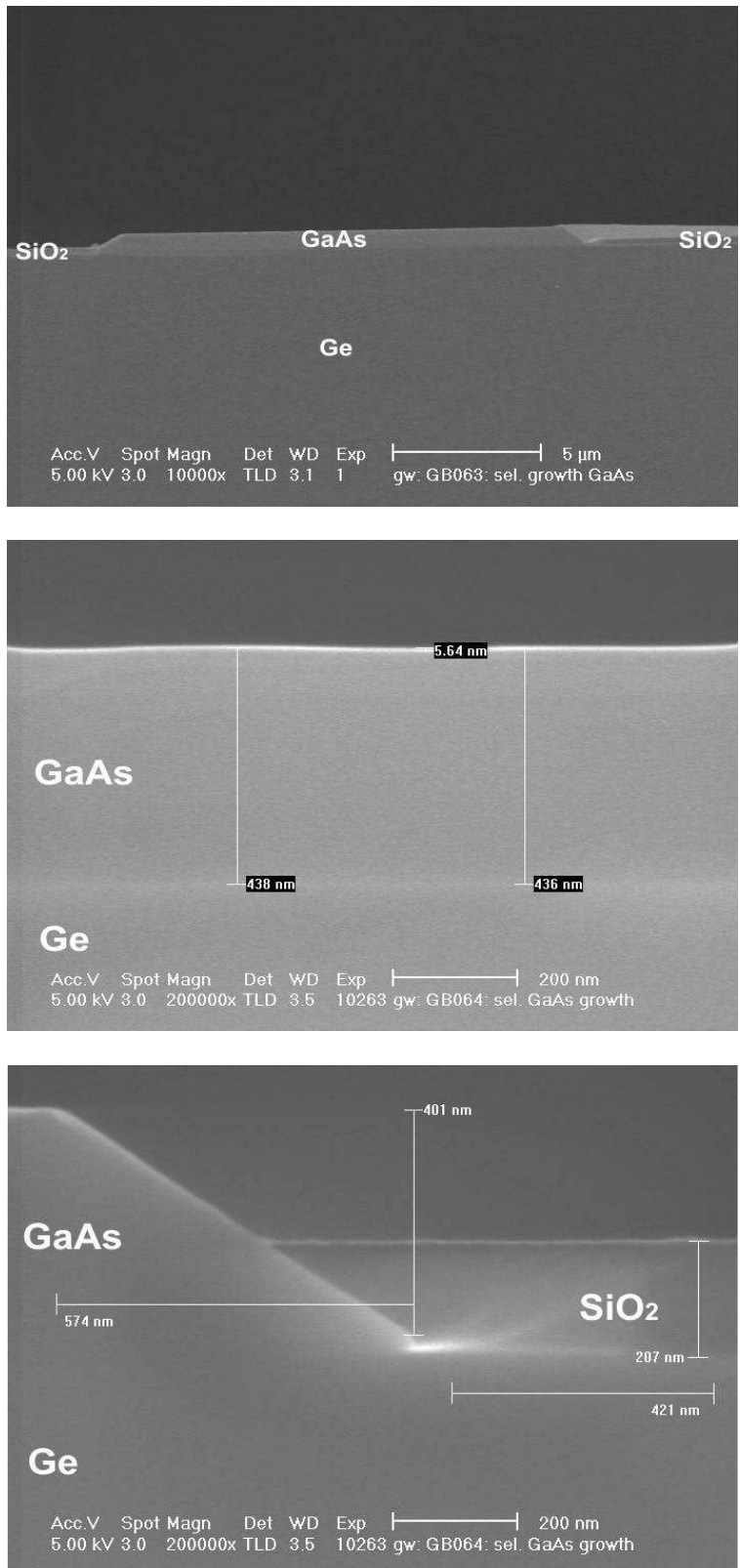

**Figure 3.** Cross section SEM images of the selectively grown GaAs layers on Ge substrates. Overview of a 12 µm wide GaAs stripe **(a)**, detail of the middle of the structure **(b)** and detail of the edge of a selectively grown GaAs structure, showing the interface with the $SiO_2$ mask layer **(c)**.

Figure 3(b) shows a detail of the bulk of the structure, showing the good flatness of the layer and revealing the thickness of the film, which is approximately equal to 440 nm. This is much thicker than the equivalent thickness of a film grown with the same growth procedure on an unpatterned substrate (150 nm) and can be explained by migration of group III precursors from the mask area to the holes in the mask. This leads to a relatively higher concentration of group III precursor in the openings and a consequently larger growth rate. The smaller the local filling factor of the mask, the larger is the effective growth rate of the selectively grown structure. Figure 3(c) shows the edge of the GaAs stripe and the interface with the $SiO_2$ mask layer. Again, the structure is inclined 45° with respect to the cleavage plane. The $SiO_2$ wet etch creates an approximately 45° slope of the edge of the $SiO_2$ layer, which can not be clearly identified on the image. The image also reveals $SiO_2$ in the background of the sample, due to the 45° angle of the structure with the cleavage plane. On the other hand, faceting of the GaAs edge and the low overgrowth over the $SiO_2$ mask can be clearly identified.

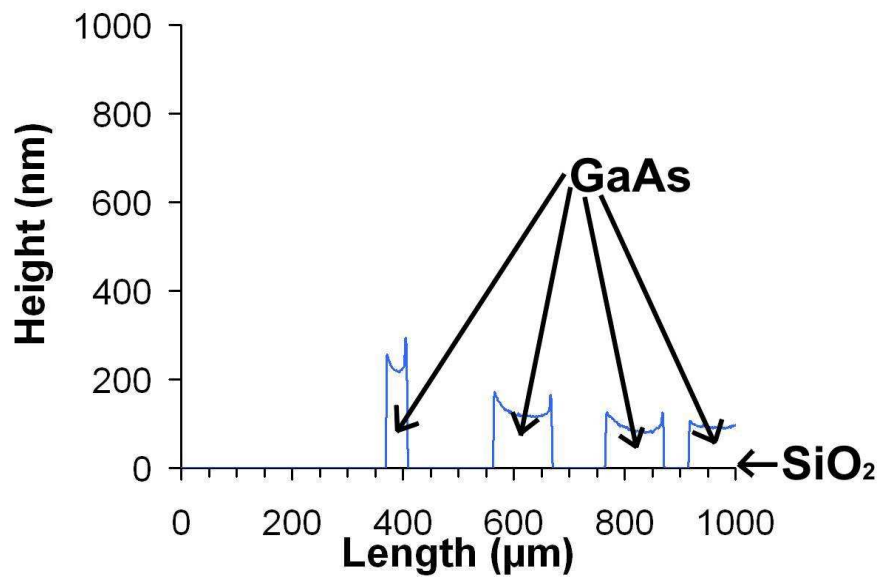

**Figure 4.** Profilometer measurement of selectively grown GaAs structures.

Figure 4 shows a profilometer measurement, which visualizes the typical U-shaped surface of the selectively grown GaAs. Also the varying heights of the different islands can be identified. Both features are a consequence of desorption of growth precursors from the mask area, redeposition in the holes of the mask and the varying local filling factor. The current growth procedure was already optimized with respect to loading effects, by providing the lowest possible growth pressure and simultaneously inhibiting the formation of APDs in the GaAs (4).

A further reduction of these loading effects can be obtained by growing in a Chlorine containing environment, either by adding HCl to the carrier gas (9,10) or by using a Cl-containing precursor (11,12). The presence of Cl in the growth environment assures the creation of very volatile and stable chlorides, increasing the desorption of group III species from the substrate and inhibiting the redeposition of these desorbed species in the openings of the mask.

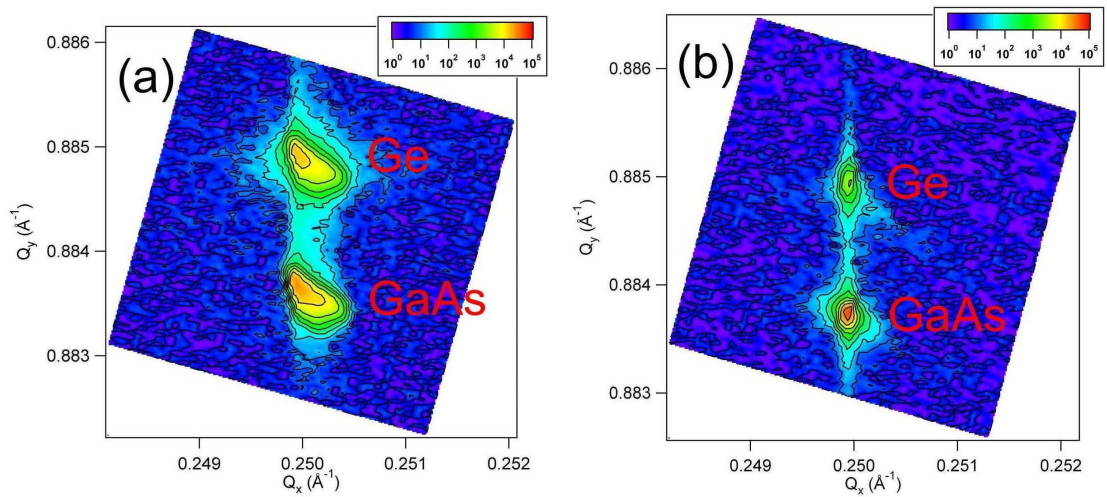

**Figure 5.** Reciprocal space map of the (115) plane x-ray reflection of a planar (a) and a selectively grown (b) GaAs film on Ge.

Having shown the macroscopic characteristics of the selectively grown layers, we now focus on the microscopic properties of the selectively grown material with x-ray diffraction (XRD) and photoluminescence (PL) spectroscopy. Figure 5 shows a reciprocal space map of the (115) reflection of a planar GaAs film and the selectively grown GaAs film on Ge. Both GaAs films are fully strained with the in-plane lattice constant of the GaAs being exactly matched to the Ge lattice constant.

A much more sensitive technique for GaAs characterization is PL spectroscopy (13). Figure 6 shows the 77 K PL spectra of the selectively grown GaAs film, as well as the PL spectrum of a planar, 1 µm thick GaAs film grown on an unpatterned Ge substrate. A one mm square area is illuminated, meaning that the intensity in the spectrum is the superposition of the PL emission of a large amount of selectively grown structures.

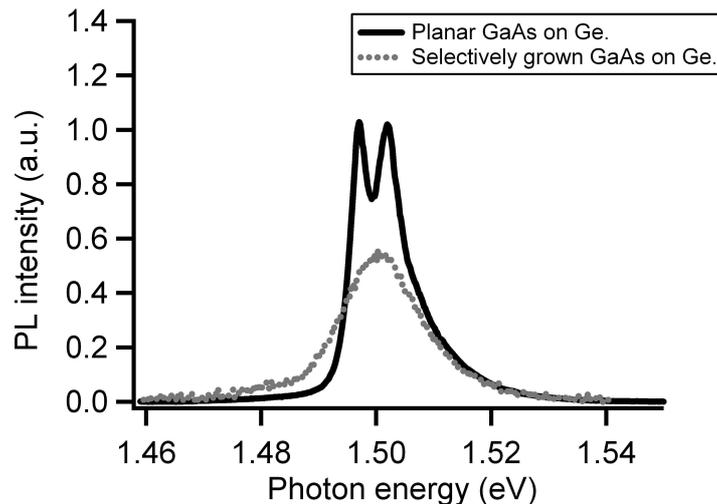

**Figure 6:** 77 K photoluminescence spectra of a selectively grown GaAs film on Ge (dotted line) and a planar GaAs film on Ge (solid line).

The PL spectrum from the high quality planar GaAs film on Ge shows two very narrow peak structures corresponding to the heavy hole and light hole band contributions, which are separated due to small strain in the GaAs layer, arising from the small 0.1% lattice mismatch between Ge and GaAs. The full width at half maximum peak width of the excitonic peaks, which translate the quality of the layer, is equal to 3 meV for the plain GaAs at 77 K, which corresponds to very high quality GaAs films. For more details about the PL structures of thin GaAs on Ge see ref. (13).

The selectively grown film shows slightly broader PL features. The heavy hole and light hole contribution can not be clearly separated, even though the shape of the spectrum suggests the presence of two overlapping peaks. Nevertheless, the 77 K FWHM energy resolution of the PL peak is still of the order of 10 to 15 meV, which corresponds to very good quality GaAs layers.

## Conclusions

We have selectively grown thin epitaxial GaAs layers on a bulk Ge substrate with a 200 nm thick $SiO_2$ mask layer. The growth is fully selective for a very large variation in filling factor over the mask area. No anti phase domain formation in the GaAs layer is observed, thanks to an optimized growth procedure that includes the growth of a 13 nm thin nucleation layer grown at relatively high pressure. Cross-section SEM and profilometer measurements reveal loading effects, both on a microscopic and a macroscopic scale. Low temperature photoluminescence measurements reveal the good quality of the selectively grown GaAs layers, presenting FWHM peak widths of the order of 10 to 15 meV.